\documentclass[preprint,showpacs,preprintnumbers,amsmath,amssymb]{revtex4}

\usepackage{graphicx}
\usepackage{dcolumn}
\usepackage{bm}

\begin{document}


\title{On The Possibility of Quasi Small-World Nanomaterials}

\author{M.A.\ Novotny and Shannon M.\ Wheeler}
\affiliation{Dept.\ of Physics and Astronomy; 
ERC Center for Computational Sciences; 
P.O.\ Box 5167; 
Mississippi State University;
Mississippi State, MS 39762-5167}
\email{man40@ra.MsState.edu}
\homepage{http://www.msstate.edu/dept/physics/profs/novotny.html}

\date{\today}

\begin{abstract}
The possibility of materials that are governed by a 
fixed point related to small world networks is discussed.  
In particular, large-scale Monte Carlo simulations are 
performed on Ising ferromagnetic models on two different 
small-world networks generated from a one-dimensional spin chain.  
One has the small-world bond strengths independent of the length, 
and exhibits a finite-temperature phase transition.  
The other has small-world bonds built from atoms, and although there is 
no finite-temperature phase transition the system shows 
a slow power-law change of the effective critical temperature of 
a finite system as a function of the system size.  
An outline of a possible synthesis route for quasi small-world nanomaterials 
is presented.  

\end{abstract}

\maketitle

\section{Introduction}

Many materials have novel physical properties that are attributed to the 
quasi-low-dimensional nature of their structure, as can be 
seen in recent research in a number of areas.  
Examples of quasi-one-dimensional materials include 
conductors and superconductors \cite{JERO87}.  
Two-dimensional thin films lead to novel effects, including 
the quantum Hall effect \cite{PELE03} and low-dimensional 
magnetism \cite{SVIS03}.  
The change from one effective dimension to another may lead to 
interesting physical effects.  For example, it may be responsible for the 
onset of high-temperature superconductivity \cite{MENZ02}.  
This type of dimensional crossover can lead to effective 
non-integer dimensional fixed points \cite{NOVO93}.  
However, quasi-low-dimensional materials have constraints due to their 
low-dimensional behavior.  These include the absence of a phase transition 
both in one-dimensional systems and in some two-dimensional systems where 
two is below the lower critical dimension.  

Recently there has been a great deal of interest in 
networks that are not regular lattices, such as 
small-world networks \cite{BARA02}.  
The study of these networks has been motivated mainly by 
social organizations (such as six degrees of separation) 
and connectivities of computers (such as the 
scale-free world-wide-web network) \cite{BARA02}.  
Such connectivities have also been used, for example, 
in non-trivial parallelization of short-ranged discrete event simulations 
\cite{KORN99,KORN00,KORN02,KOLA03,KORN03}.  
Simulations of models such as Ising spin models on these networks have been 
studied, but no attempt has been made to ask whether these 
theoretical models could actually be designed and built via 
various synthesis routes.  This question is addressed 
in this manuscript.  There is a difference between the models studied to 
date and the question of whether or not materials can be made 
that are governed by small-world fixed-points.  
The difference is that materials 
must be constructed from atoms and must be embedded in three-dimensional 
space.  

\section{Model and Methods}

The models studied here are Ising models with 
$N_0$ spins on a linear chain, with a nearest-neighbor 
ferromagnetic interaction constant $J_1$.  

In the first model, model~1, if 
there is a small-world connection between spins $i$ and $j$ 
and $i\ge j+1$ 
a ferromagnetic interaction of strength $J_2(i,j)$ is added.  
The Hamiltonian is 
\begin{equation}
\label{eqHAM1}
{\cal H} = - J_1 \sum_i \sigma_i\sigma_{i+1} - 
\sum_{\rm SW} J_2(i,j) \sigma_i \sigma_j
\end{equation}
where periodic boundary conditions are used (spin $N_0+1$ equals spin 1)
and the Ising spins $\sigma=\pm1$.  Terms in the second sum 
are only present if there exists a small world connection between 
spins $i$ and $j$.  We construct a (random) small-world network 
algorithmically by: i) start with spin 1, and 
randomly connect it to any of the other $N_0-1$ spins; 
ii) if spin 2 is not connected to spin 1, then randomly connect it to 
one of the $N_0-2$ spins that are not already connected; iii) continue 
for all $N_0$ spins.  Note that here $N_0$ must be even.  
This algorithm gives each spin $i$ three connections, 
two of strength $J_1$ to spins $i-1$ and $i+1$ and one of strength $J_2$ to 
the small-world connection between spin $i$ and $j$.  
This type of small-world connection has been used in the past to 
obtain perfectly scalable parallel discrete-event simulations \cite{KORN03}.  
A study of the Ising model on a similar small world 
network with $J_2(i,j)=J r^{-\alpha}_{ij}$ (i.e.\ having a power-law 
dependence on distance) has recently been performed \cite{JEON03}.  
There they conclude that any non-zero value for $\alpha$ 
destroys the finite-temperature phase transition in 
the thermodynamic limit.  
This result differs from the case with $J_2(i,j)=J_2$ (independent 
of length), where a finite-temperature phase transition has 
been observed \cite{GITT00,BARR00,PEKA01,HONG02,KIM01}.  
Our first model has all small-world bonds of strength $J_2$, 
independent of length.  

In our second model, model~2, we construct the small-world connections 
from Ising spins.  In particular, for each small-world bond 
constructed as in our first model between spins $i$ and $j$, we add 
$r_{ij}+1$ Ising spins and couple each of them together with 
interactions of strength $J_2$.  These additional Ising chains are coupled 
to the original Ising spins $i$ and $j$ with interaction strength
$J_2$.  Thus, the small-world connections have been constructed using 
the same lattice constant as the original lattice (but different 
nearest-neighbor interactions).  

We have performed a standard importance-sampling Monte Carlo 
simulation \cite{LAND00}.  
We measured the magnetization $m=\sum_i\sigma_i/N_{\rm tot}$, 
order parameter, 
susceptibility $\chi$, internal energy, specific heat, and the Binder 
4$^{\rm th}$ order cumulant for the order parameter 
$U=1-
\frac{\left\langle m^4\right\rangle}{3\left\langle m^2\right\rangle^2}$.  
The code was run using trivial parallelization on a cluster, using up to 
128 processing elements.  The parallel random number generator 
SPRNG 1.0 \cite{SPRNG} was used.  The spins to be updated were chosen 
randomly, a Glauber spin-flip probability was used.  
For each temperature $T$, $10^5$ Monte Carlo steps per spin (MCSS) 
were used for thermalization and averages were taken over 
$10^6$ MCSS (with measurements taken every $10$ MCSS).  
The energy units were chosen such that $J_1/k_{\rm B}=1$, 
where $k_{\rm B}$ is Boltzmann's constant.  
In this paper the choice $J_2=4J_1$ was made.  

\begin{center}
\begin{figure}
\includegraphics[width=.75\textwidth, bb= 38 45 534 455]{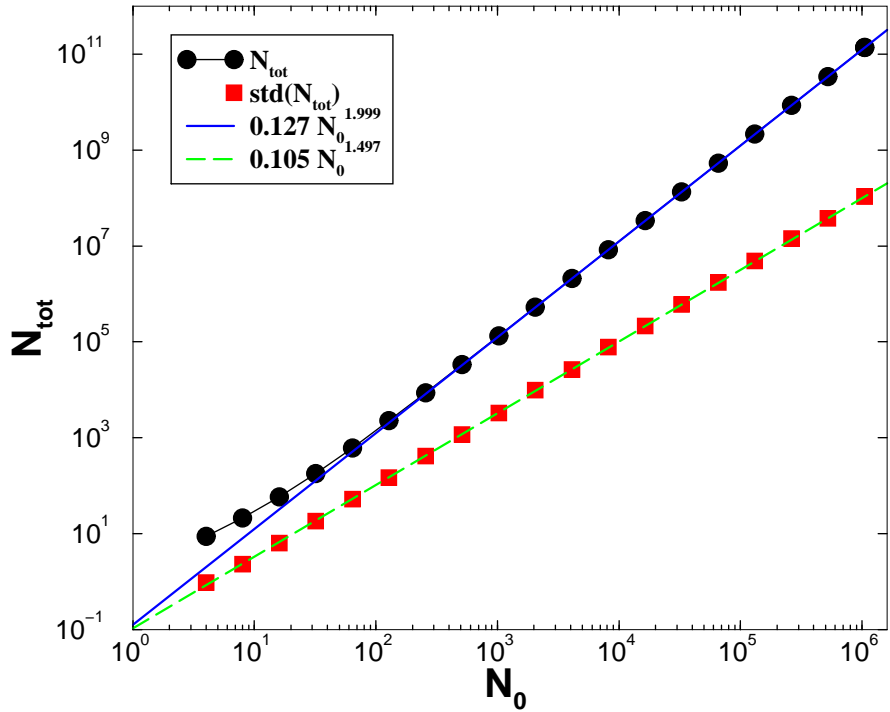}
\caption[]
{The total number of Ising spins $N_{\rm tot}$ as a function 
of the number of spins $N_0$ in the one-dimensional 
chain is shown.  The solid line is a fit to the data for $N_0\ge2^{10}$.  
Also shown is the standard deviation of $N_{\rm tot}$ from the mean 
(squares), and a fit to those data (dashed line).
}
\label{Fig1}
\end{figure}
\end{center}

\section{Data and Analysis}

In our second model, one question is the average number of total 
Ising spins $N_{\rm tot}$ starting with a spin chain of length $N_0$.  
For a chain of length $L$ and lattice constant $a$, one has $L=N_0 a$.
Here we take the lattice spacing $a=1$, so $N_0=L$.  
It is shown in Fig.~1 that 
$N_{\rm tot}\propto N_0^x$ with exponent $x\approx 2$.  A fit 
to data (averaged over $10^3$ different small-world bond connections) 
for $2^{10}\le N_0\le2^{20}$ gives 
$N_{\rm tot} = 0.127 N_0^{1.999}$.  
A fit with the same range of $N_0$ for 
the standard deviation of the mean gives 
${\rm std}(N_{\rm tot})=0.105 N_0^{1.497}$.  

The relationship $N_{\rm tot}\propto N_0^2$ can be understood easily 
in an approximate fashion, by ignoring correlations between the 
small-world interconnections.  
Although the number of spins scales as the square of the linear 
distance $N_0$, the fixed point governing the system 
is not expected to be the two-dimensional fixed point since the 
bond connectivity is different from that of a two-dimensional lattice.  
The average length of a connection is $N_0 a/4$.  This is because the 
lengths of the connections are chosen uniformly up to the maximum 
possible distance of a connection, which is $N_0 a/2$ due to the 
periodic boundary conditions.  
There are $N_0/2$ such connections (the factor of 2 is to take into account 
double counting).  Ignoring factors in the number of spins that are 
proportional to $N_0$ thus gives $N_{\rm tot}\approx \frac{1}{8} N_0^2$.
As seen in the fit in Fig.~1, both the fitted 
exponent and prefactor are in excellent agreement with this argument 
for $N_0\ge2^7$.  
An extension of this argument shows that if 
only a fraction $p_{\rm connect}$ of spins 
were connected, then the total number of spins should 
scale like $N_{\rm tot}\approx \frac{p_{\rm connect}}{8}N_0^2$ for large 
$N_0$.  Simulations were performed with values of 
$N_0=8$ and for $N_0$ up to $N_0=16384$ for model 1 and 
$N_0=256$ for model 2.  

\begin{center}
\begin{figure}
\includegraphics[width=.48\textwidth, bb= 38 45 534 455]{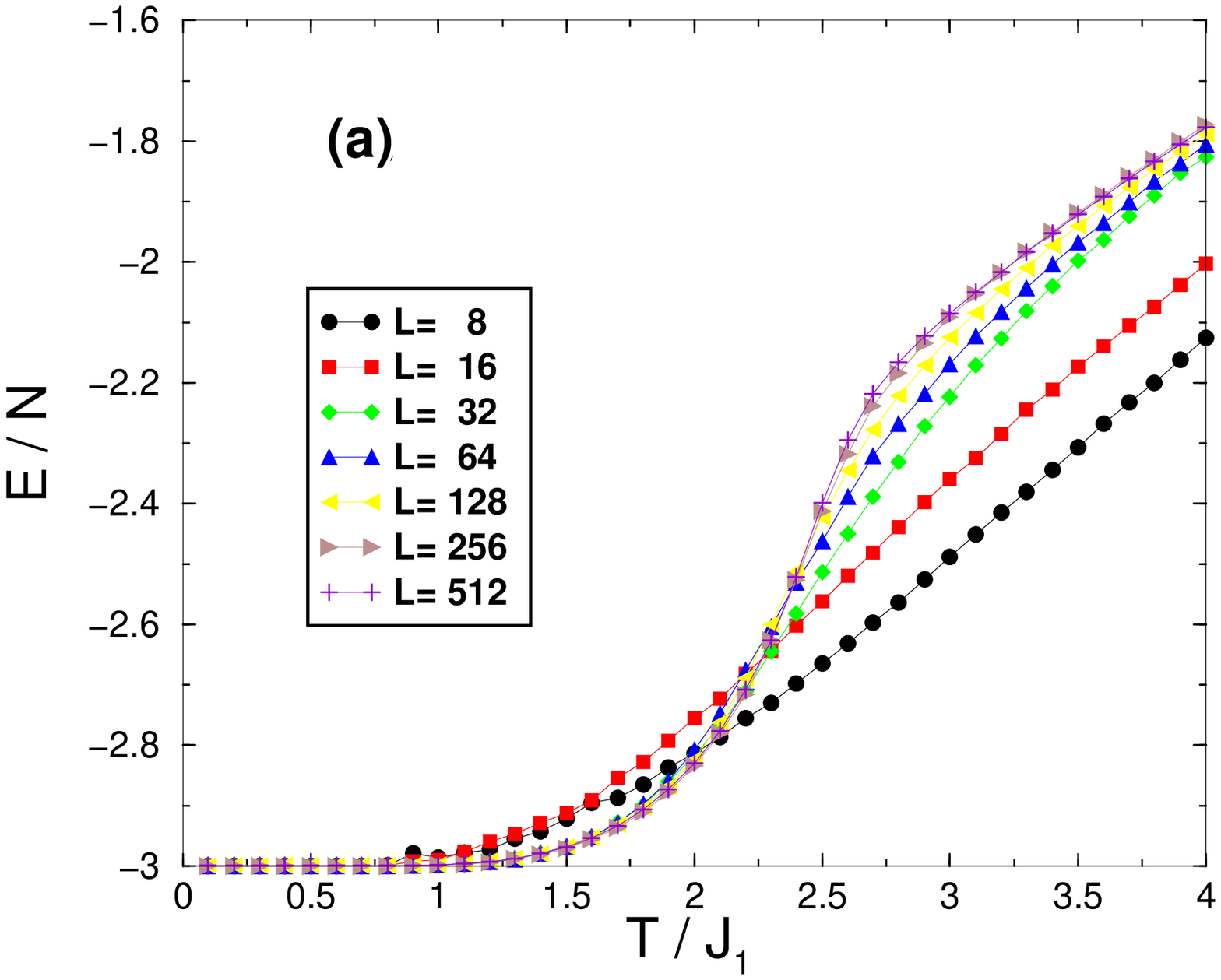}
\includegraphics[width=.48\textwidth, bb= 38 45 534 455]{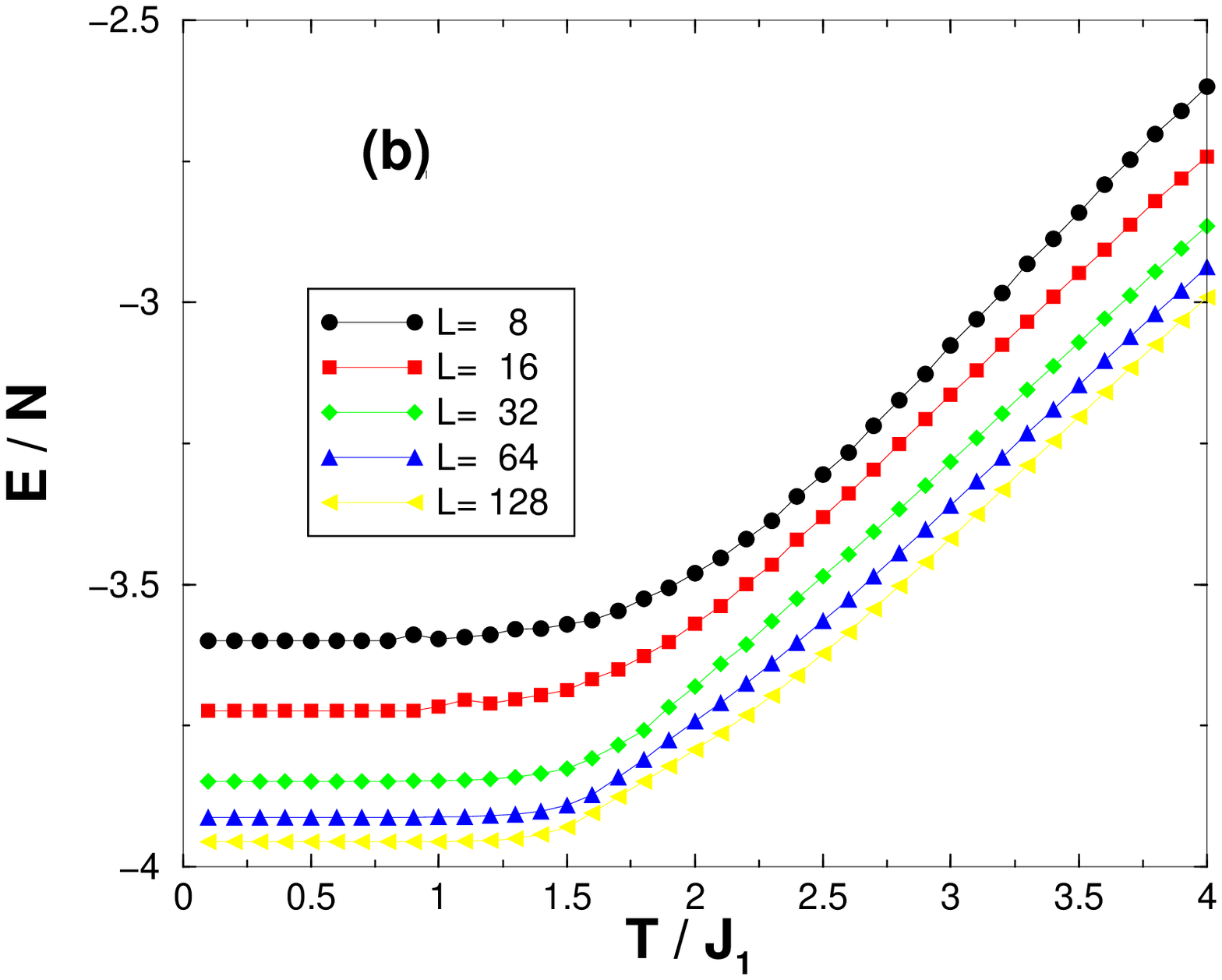}
\caption[]
{The internal energy per Ising spin for (a) model 1, and (b) model 2.
These are parameterized by the chain length, $L=N_0 a$, where $a$ 
is the lattice spacing and $N_0$ is the number of spins in the 
one-dimensional backbone.  
}
\label{FigE}
\end{figure}
\end{center}

Figure~\ref{FigE} shows the average internal energy per Ising spin, 
in units of $J_1$, 
as a function of temperature.  The internal energy is the 
expectation value of the Hamiltonian, $E=\langle{\cal H}\rangle$.
Model~1, Fig.~\ref{FigE}(a), 
has the low-energy value for $E/N$ independent of $N$, since 
in the ground state (all spins the same direction) the energy 
is $E/N=(2 J_1 +J_2)/2$ where the division by $2$ takes into account 
double counting.  For model~1, $J_1=1$ and $J_2=4$, so $E/N=-3$ at 
low temperatures.  Note that model~1 has a singularity 
developing with large $N$ near $T=2.5J_1$.  
Model~2, Fig.~\ref{FigE}(b), looks different.  
Note that for model~2, this energy is 
divided by the total number of Ising spins, $N_{\rm tot}$.  
The ground-state energy per spin has the form
$\frac{E}{N_{\rm tot}}=\frac{N_0\left(2J_1+J_2\right) + 
2J_2\left(N_{\rm tot}-N_0\right)}{2 N_{\rm tot}}$, which 
at low temperatures approaches $-4$ since $N_{\rm tot}\gg N_0$ for 
large $N_0$.  

\begin{center}
\begin{figure}
\includegraphics[width=.48\textwidth, bb= 38 45 534 455]{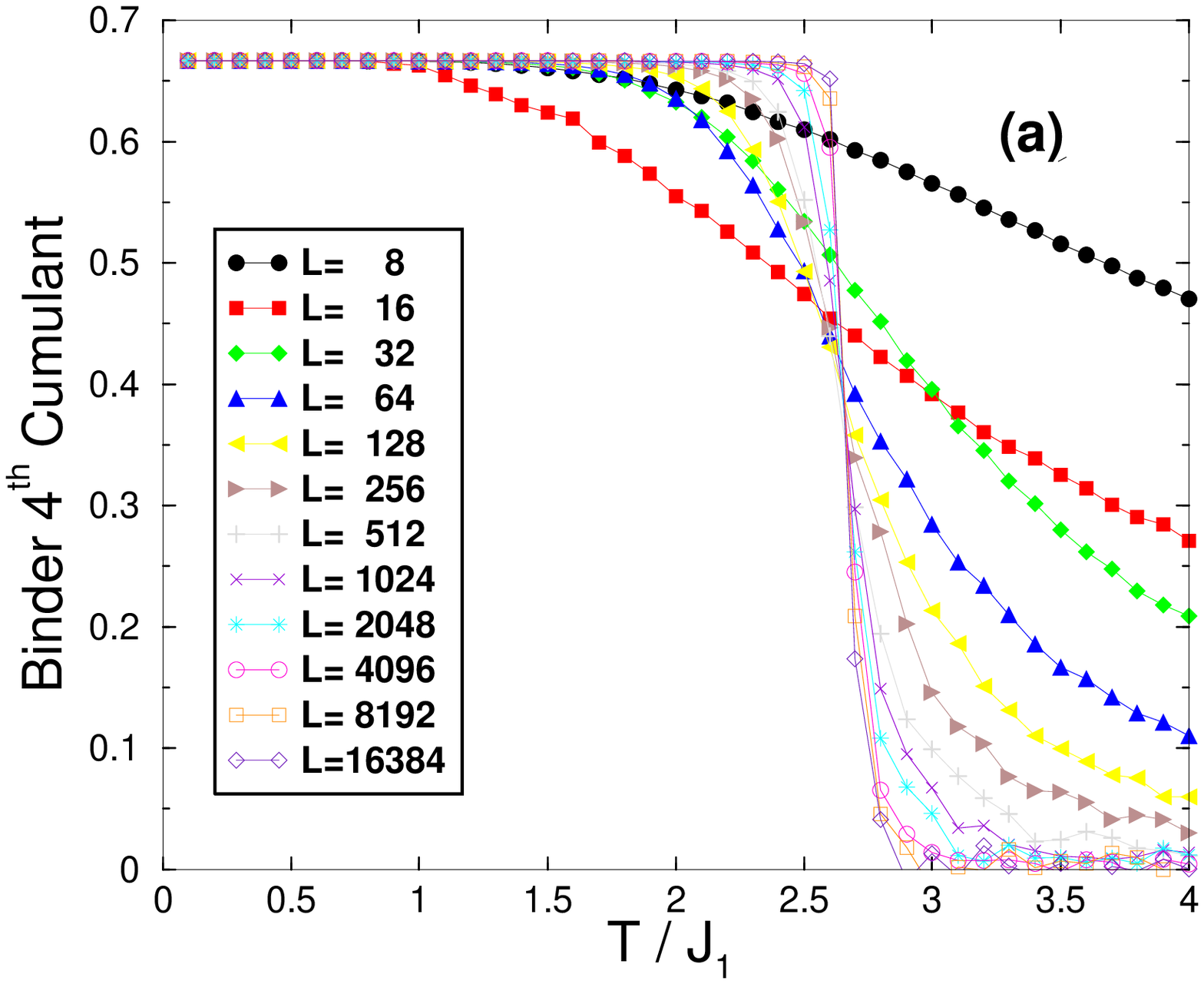}
\includegraphics[width=.48\textwidth, bb= 38 45 534 455]{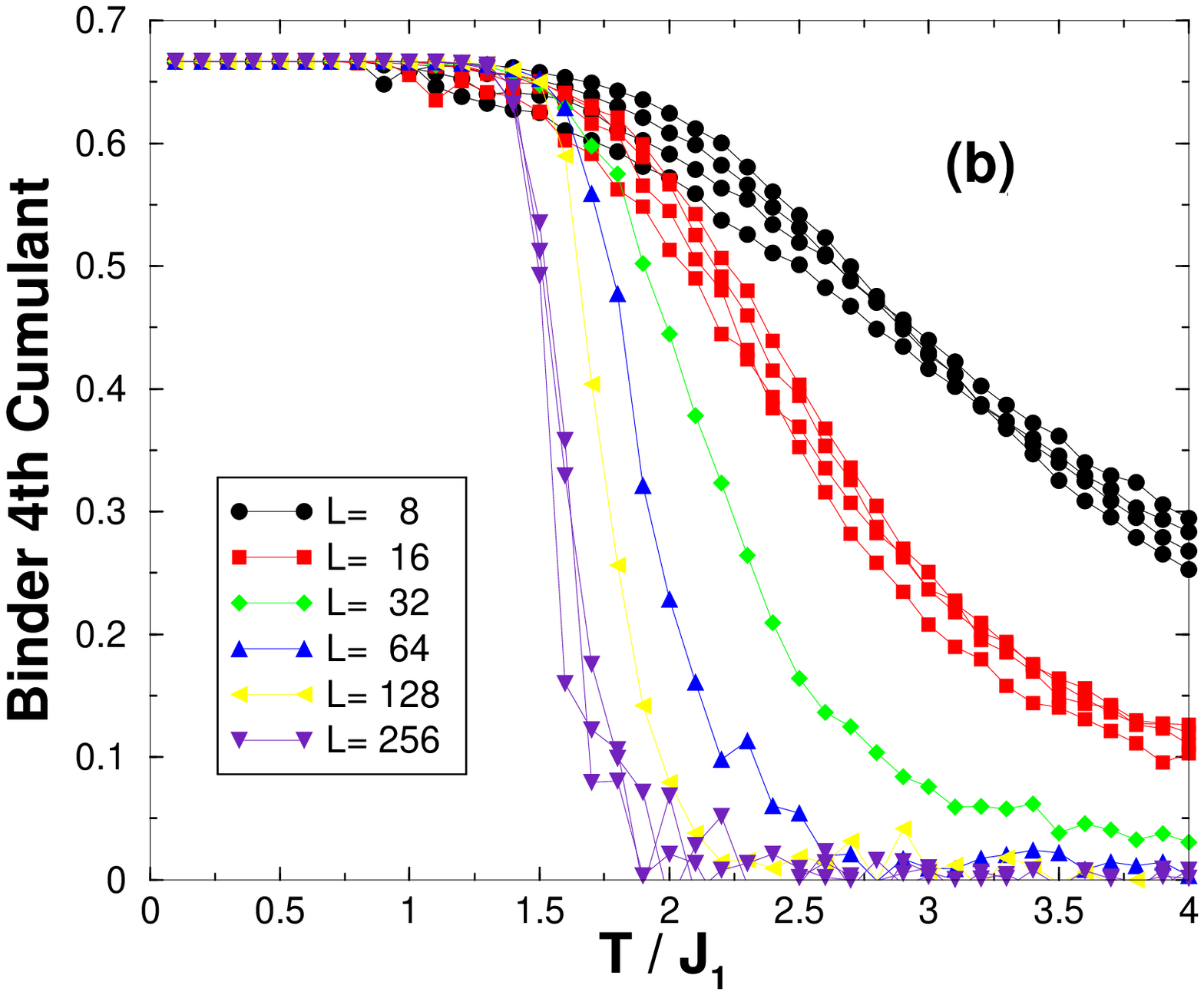}
\caption[]
{The Binder fourth-order cumulant for (a) model 1, and (b) model 2.
}
\label{FigBinder}
\end{figure}
\end{center}

Figure~\ref{FigBinder} shows the Binder $4^{th}$-order cumulant for 
the magnetization as a function of the temperature for both models.  
The temperature where two of these curves for different $N_0$ 
cross is an estimate for the critical temperature.  An alternative 
estimate can be obtained by choosing a value such as 0.5, and 
letting the estimate of the critical temperature be where the 
cumulant crosses the value of 0.5.  (Ideally, one would like to 
use the value of the infinite-lattice cumulant, but this is 
often unknown unless the universality class of the model is 
known.)  
Model~1, Fig.~\ref{FigBinder}(a), shows a very nice indication of 
a critical point near $T\approx 2.7 J_1$.  
Model~2, Fig.~\ref{FigBinder}(b), shows no such estimate of the 
critical point.  Only by choosing the temperature where the 
cumulant crosses 0.5 can an estimate of the finite-size 
critical point (if it exists) be obtained.  
Note that for model~2 there are a number of different quenched 
small-world bond configurations for $N_0=8$, $16$, and $256$.

\begin{center}
\begin{figure}
\includegraphics[width=.48\textwidth, bb= 38 45 534 455]{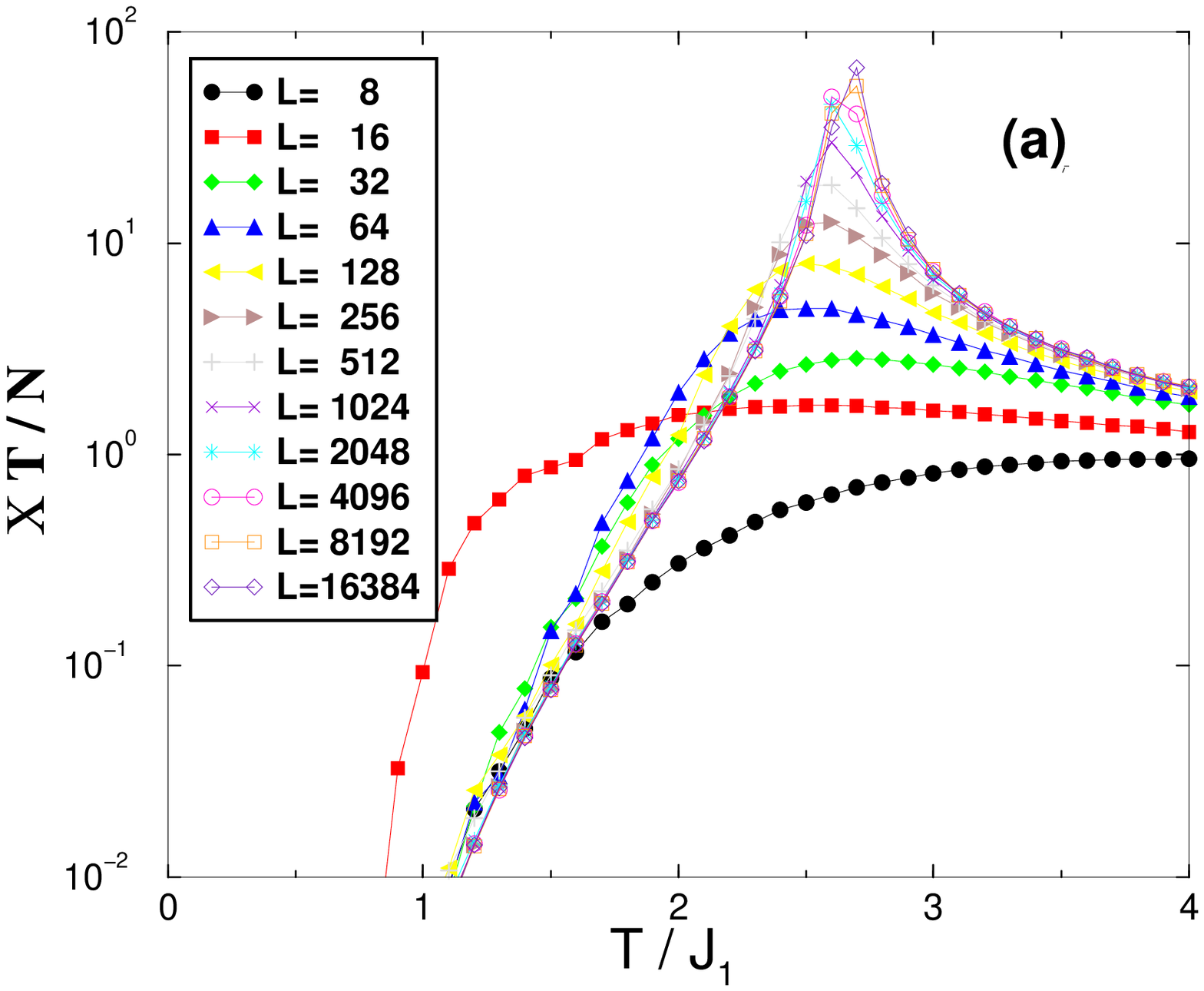}
\includegraphics[width=.48\textwidth, bb= 38 45 534 455]{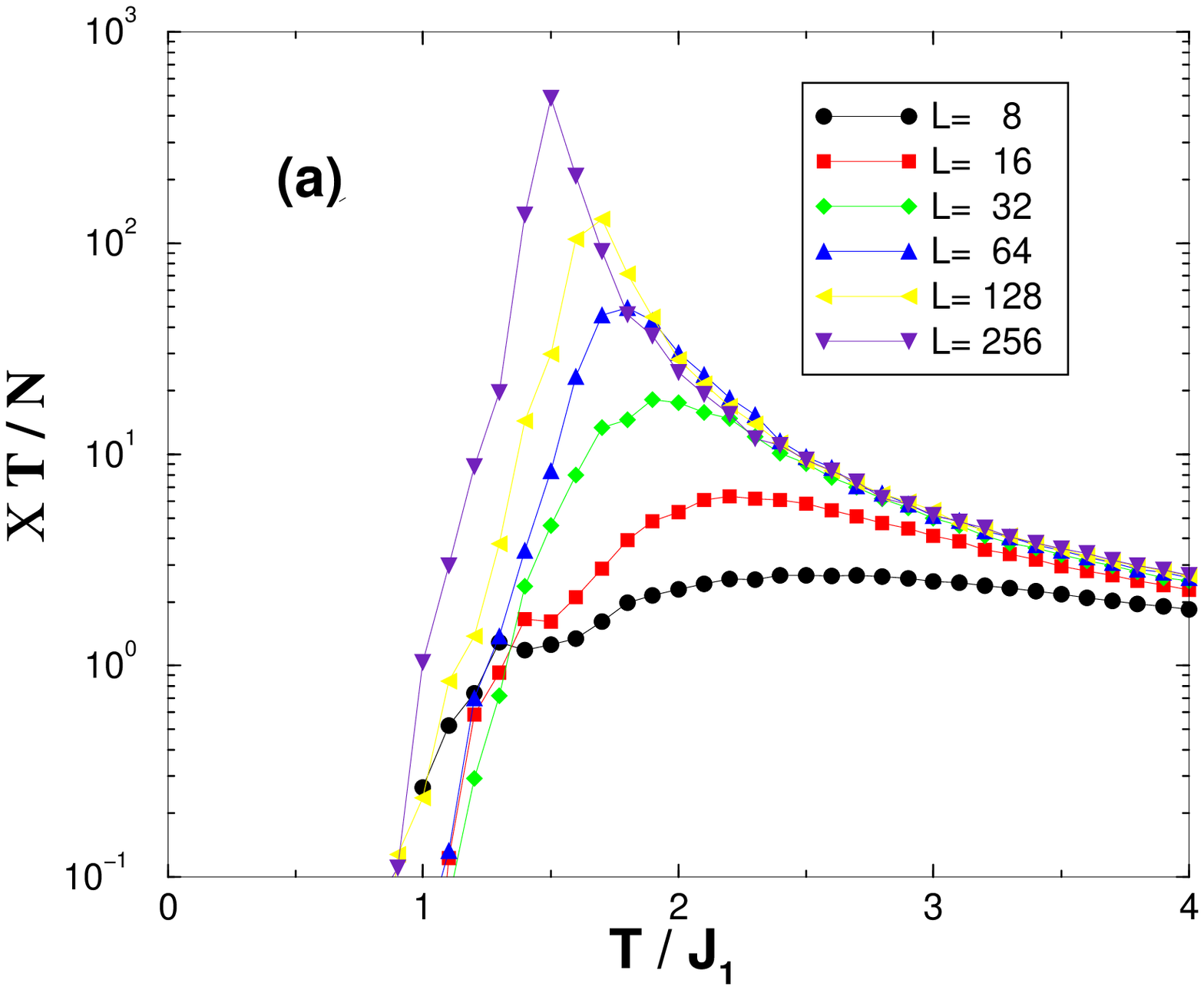}
\caption[]
{The susceptibility times temperature for (a) model 1, and (b) model 2.
}
\label{FigChiT}
\end{figure}
\end{center}

\begin{center}
\begin{figure}
\includegraphics[width=.75\textwidth, bb= 38 45 534 455]{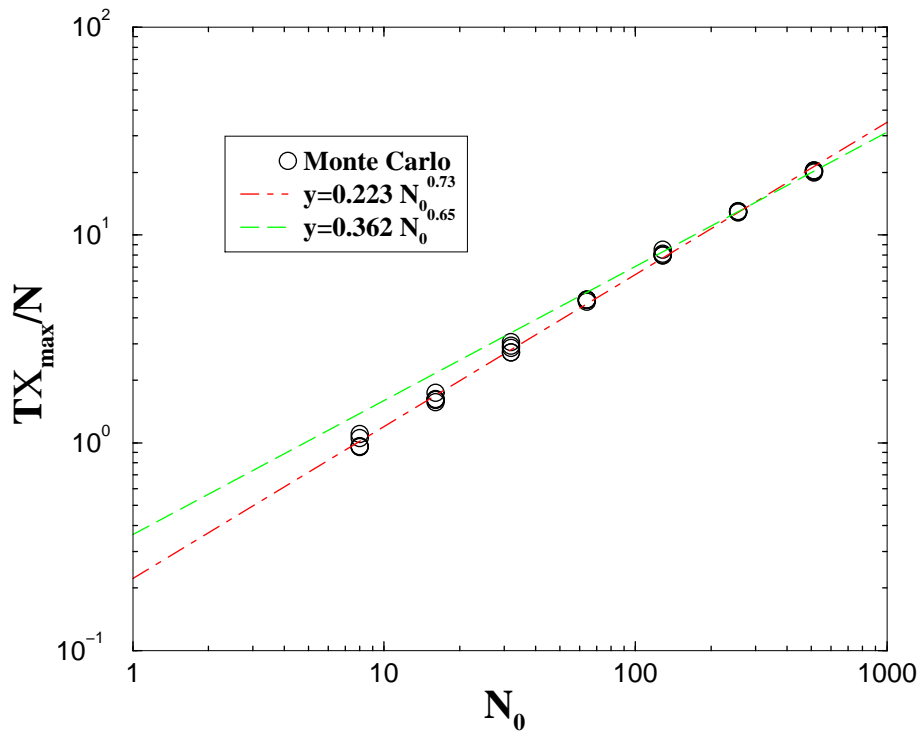}
\caption[]
{The maximum of the susceptibility times the temperature for model~1.
The slope of this gives an estimate for the ratio of critical 
exponents $\gamma/\nu$.  
}
\label{FigChiTmax}
\end{figure}
\end{center}

Figure~\ref{FigChiT} shows the susceptibility $\chi$ per Ising spin times the 
temperature $T$.  
For model~1, Fig.~\ref{FigChiT}(a), shows a sharpening maximum 
at $T\approx 2.7$.  Figure~\ref{FigChiTmax}, for 
model~1, shows the maximum value for $\chi T$ as a function of $N_0$.  
There are five different quenched small-world connections shown for 
each value of $N_0$.  
The slope of this log-log plot gives an estimate for 
the exponent ratio $\gamma/\nu$, with $\gamma$ the 
susceptibility exponent and $\nu$ the correlation length exponent.  
The value obtained for the slope is $0.73$ using all lattice 
sizes, and it is $0.65$ using only the sizes $256$ and $512$.  
Although simulations for larger $N_0$ may give lower slopes yet, 
we estimate that $\gamma/\nu=0.6\pm0.1$.  

\begin{center}
\begin{figure}
\includegraphics[width=.75\textwidth, bb= 38 45 534 455]{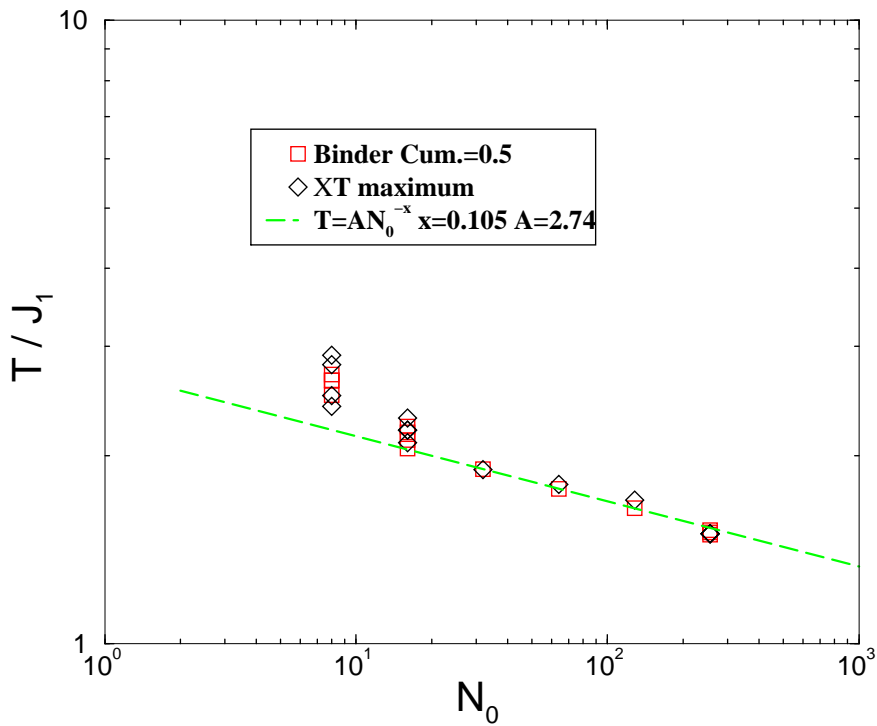}
\caption[]
{The summary for model 2 of the various indicators of critical behavior 
are shown as a function of the system size.  Note that this fit 
(dashed line) suggests 
that a finite-temperature phase transition does not survive 
taking the thermodynamic limit.  
}
\label{Summ1}
\end{figure}
\end{center}

The maxima of $\chi T/N_{\rm tot}$ 
for model~2, Fig.~\ref{FigChiT}(b), decrease with 
temperature as $N_0$ increases.  This indicates that there may be 
no finite-temperature phase transition in the thermodynamic limit.  
This conclusion is further supported in Fig.~\ref{Summ1}.  
There the location of both the crossing of the Binder cumulant and 
0.5 and the maximum of $\chi T/N_{\rm tot}$ is shown.  Both estimates 
for the `finite system-size' critical point agree.  
They both also have a reasonable fit, based on $N_0\ge 32$, of 
$T_{\rm c}=A N_0^x$ with $x=-0.105$ and $A=2.74$.  

\section{Possibilities for Small World Materials}

Model~2, with the small-world bonds built of atomic spins, 
exhibits no finite temperature phase transition in the 
thermodynamic limit.  Consequently, the outlook for 
small-world materials is that they will not be easy to synthesize.  
Note that rigorously it has been shown that 
no finite temperature phase transition can occur in one dimension.  
Nevertheless, 
there are a large number of effective one-dimensional 
materials \cite{JERO87}. Small-world models with fixed 
connection sizes, such as model~1, exhibit a 
finite temperature critical point.  
Building these small-world connections from atomic spins as in model~2 
provides the small-world bonds.  However although the small-world 
fixed point is unstable at finite temperatures, 
the flow from this unstable fixed point seems to be extremely slow.  
Consequently, the outlook for quasi small-world materials is better than 
for quasi-one-dimensional materials.  

Table~I shows the effective $T_{\rm c,eff}$ for a given size of 
material for model~2.  Also shown from ref.~\cite{JEON03} are values 
of $T_{\rm c,eff}$ for small-world connections with strengths that 
behave as $J_1 r_{ij}^{-\alpha}$, for different values of $\alpha$.  
All temperatures are in units of $J_1$.  Due to the 
slow decrease in the critical point with the system size, 
Table~I shows that even for large systems (of the size of meters) the 
system should exhibit an effective critical temperature that can 
still occur at a reasonably high temperature.  

\begin{table}[!t]
\begin{tabular}{|c|r||r||r|r|r|}
\hline
Material & $N_{\rm atom}$ & $T_{\rm c}$ & $T_p$ & $T_p$ & $T_p$ \\
Size     & & added atoms & $\alpha=0.1$ & $\alpha=0.2$ & $\alpha=0.3$ \\
\hline
{\it base} & $1$ & $2.74$ & $1.62$ & $1.67$ & $1.69$ \\
\hline
$\mu$m & $5\times10^3$ & $1.12$ & $1.48$ & $1.31$ & $1.69$ \\
\hline
mm & $5\times10^6$ & $0.54$ & $1.37$ & $1.02$ & $0.70$ \\
\hline
m & $5\times10^9$ & $0.26$ & $1.26$ & $0.73$ & $0.26$ \\
\hline
\hline
\end{tabular}
\caption{Extrapolated values for the 
finite-size effective critical temperature (for example $T_p$, 
the location of the maximum in $\chi T$) for various 
sizes corresponding to the given number of 
atoms in a one-dimensional chain material.  
See the text for the assumptions.  The last three columns 
represent the data of ref.~\cite{JEON03}, as explained in the text.  
}
\end{table}

One synthesis route for small-world nanomaterials is described here.  
This is a theorist's cartoon of the synthesis route, and 
consequently is not meant to be detailed.  First, start by constructing 
a one-dimensional chain of atoms (or molecules), 
most likely embedded on the surface of a support material.  
Prepare a solution of linear molecules of all different lengths up 
to the maximum size of material to be synthesized.  This could be 
accomplished by starting with linear molecules of the same length, 
and then having a chemical process that breaks the linear chains 
at random locations.  Place a reactive agent at the ends of these 
linear segments, this agent should be reactive with the 
constructed one-dimensional chain of atoms on the surface of the material.  
Bring this constructed chain of atoms into contact with the 
solution of reactive linear molecules, forming the 
small-world connections.  A further synthesis step can be performed 
to reduce the length of the linear molecules to the minimum possible 
distance.  This could be accomplished by removing 
atomic segments.  The 
material synthesized in this way should mimic closely the 
small-world connections of model~2.  

There are several factors that may make quasi small-world nanomaterials 
easier to synthesis than the synthesis route outlined above. 
It has been shown that not all 
original atoms (ones that are not in the small-world connections) 
need to be connected with 
small-world bonds for the system to be controlled by the 
small-world fixed point \cite{KORN03,JEON03}.  
Rather, what is required is just a finite density of small-world bonds.  
Hence, the prefactor in front of the 
power law for the number of total atoms per original atom may be made 
arbitrarily small. It is also possible to consider starting with a 
two-dimensional thin film, and synthesizing small-world connections 
on the film in a fashion similar to the one-dimensional route outlined 
above.  Some studies of crossover from 
two- and three-dimensional fixed points to small-world fixed points 
have been performed \cite{HERR02,HAST03}.  However, it is anticipated 
that for any finite density of small-world connections the dominant 
fixed point should be the small-world fixed point.  
Furthermore, since these original one and two-dimensional 
systems are embedded in three dimensions, it is possible to 
make some of the longer-distance small-world connections 
very short-ranged by bending or folding the original lattice 
to minimize the total length of all small-world bond connections.  
Finally, just as in quasi-one-dimensional materials there is the 
possibility of allowing weak interactions in three dimensional crystals 
to stabilize the fixed point in different dimensions before the 
system exhibits a cross-over to the three-dimensional fixed point.  

\section{Discussion and Conclusions}

The ferromagnetic Ising model on two 
small-world networks has been studied. 
One model, related to models previously studied by others 
\cite{GITT00,BARR00,PEKA01,HONG02,KIM01}, 
has only fixed-strength interactions for the small-world connections.  
This model exhibits a finite-temperature second-order phase transition.  
We have determined from finite-size scaling of the maximum of the 
susceptibility the exponent ratio $\gamma/\nu= 0.6(1)$ for the 
system sizes simulated.  
A study by other researchers 
with small-world connections that fall off as a power-law with 
actual distance shows no finite-temperature phase transition, but a 
logarithmic decrease with $N_0$ of an effective finite system size critical 
temperature $T_{\rm c,eff}$\cite{JEON03}.  

The second Ising model studied here 
is more realistic for the possibilities of 
quasi small-world materials.  The small-world connections have been 
constrained to be built from spins using the same lattice spacing.  
This models what would have to be accomplished in actual quasi small-world 
materials, since atoms are the fundamental building blocks.  
For this model, no finite-temperature phase transition was found 
to survive the thermodynamic limit.  In particular, 
in terms of the number of linear spins $N_0$, an effective 
critical temperature $T_{\rm c,eff}$ was found to 
behave as $N_0^x$ with $x\approx-0.105$.  Furthermore, on 
general grounds the total number of spins scales 
as $N_{\rm tot}\sim N_0^2$ for a one-dimensional 
system with $N_0$ spins.  

From these studies a theoretical picture emerges about the difficulty 
in synthesizing quasi small-world materials.  The number of 
atoms needed to connect small-world bonds may be large, and is 
expected to scale as a power law of the number of atoms in the 
none small-world system.  However, it has been shown that not all 
none small-world atoms need to be connected with small-world bonds 
for the system to be controlled by the small-world fixed point 
\cite{KORN03,JEON03}.  Consequently, the prefactor in front of the 
power law for the number of small-world atoms may be made 
arbitrarily small in principle.  
For strengths of small-world connections that decrease as a 
power law \cite{JEON03}, a logarithmic fall-off of the 
effective critical temperature with system size has been found.  
Using some order-of-magnitude assumptions and the different power law results 
from this work and ref.~\cite{JEON03}, one finds that 
a reasonable value of 
$T_{\rm c,eff}$ is possible for systems in the size ranges of 
microns, millimeters, or even meters.  
For the model studied here, where the small-world connections were 
built from atomic spins, a small power-law fall-off of the effective 
critical temperature was found.  Thus nanoscale and mesoscale 
materials should exhibit an effective critical behavior related to 
the small-world bonds.  A possible synthesis route to materials 
effectively governed by small-world fixed points was outlined.  

\section{Acknowledgements}

I acknowldge useful discussions with a number of people, particularly 
Torsten Clay, 
Seong-Gon Kim, 
Alice Kolakowska, 
Gyorgy Korniss, 
Chris Landee, 
Per Arne Rikvold, 
and 
Zoltan Toroczkai.  
Supported in part by NSF grants DMR-0120310 and DMR-0113049.  
Computer time from the Mississippi State University ERC and 
NERSC was critical to this study.

\end{document}